\documentstyle[12pt]{article}

\begin{document}

\title{ \begin{flushright}
{\large \bf IC/94/155}
\end{flushright}
\Large \bf VALENCIA 93: THE SUMMARY OF PARTICLE THEORY\\}
\author{ {\large \bf
 Goran Senjanovi\'c }\\
\it International Centre for Theoretical Physics,\\
\it Trieste, Italy\\
E-mail: goran@ictp.trieste.it}

\maketitle

\abstract{ The International School on Cosmological Dark Matter held
in Valencia in the
fall of 1993 was devoted to the
interplay of cosmology and particle physics, with the obvious emphasis on the
Dark Matter issue. Here I present the expanded version of my summary talk
regarding the particle physics theory part of the School. }

\newpage

\section{Introduction}

I am deeply aware that it is impossible to give a complete summary of a week
long meeting, but I will still try, if for no other reason but for the
months long push by the organizers. My task has been made easier by the
fact that the summary of the cosmology part of the School has been done by
Gary Steigman, although I still expect to fail.

My plan is the following. I will not discuss talks one by one, but rather
 go over what in my opinion were the highlights of the meeting, and
what  the important open problems of present day particle physics are.
Needless to say my presentation will be subjective, if not biased, and I
apologize to those speakers who may find their contributions ignored or
not well presented. It will not happen for reasons of ignorance and bias
only, but at least partially for the lack of time and space.

The theory part of the school started with my lectures on the standard
model of particle physics. My enthusiasm in accepting this was highly
motivated by Jose Valle's twisting my arm, although I should mention that he
was nice enough to let me go, in his own words, {\em un poquito} beyond. I
will concentrate here mostly on the {\em poquito} part, but  for  the sake of
completeness we shall start with the Standard Model and its relation with
cosmology. The second part of my review will then be devoted to the
interesting extensions of the Standard Model, always keeping in
 mind cosmology.

\section{The Standard Model and \newline Cosmology}

For the sake of clarity and in order to establish the notation and the
conventions, let me define the Standard Model (SM) as the
$SU(2)_L\otimes U(1)$ gauge theory of electroweak interactions with three
families and one Higgs doublet.

We all know that the Standard Model is in excellent agreement with the
experiment, far better than any of us would have imagined  twenty years ago.
There is not even a shred of evidence of any new phenomena beyond the
electroweak scale $M_W$. It is then natural that we should turn to
astrophysics and cosmology in the attempt to uncover some new physics.
There are many such issues, but I will concentrate on the few main ones
only. As the name of the School dictates, we will start with the problem
of the Dark Matter (DM), and go to the solar neutrino puzzle (SNP),
baryogenesis, and then only briefly mention the problem of atmospheric
neutrinos (AN).

Although phenomenologically extremely successful, the Standard Model
suffers from two serious theoretical problems that encourage
 the search for a more complete theory.

\begin{enumerate}

\item {\bf The problem of Higgs mass}. The scalar
mass terms are not protected by any symmetry and so it seems that there
is no mechanism that can prevent the Higgs particle from acquiring
arbitrarily large mass. In other words, how do we keep $M_W \ll M_{Pl}$
?
\item {\bf The strong CP problem}. In the Standard Model the effective strong
CP parameter
\begin{equation}
\bar\theta = arg(F\tilde F\; term)_{QCD} + arg ( det M_q)
\label{theta}
\end{equation}

($M_q$ being the quark mass matrix) is not controlled in perturbation
theory.  It gets infinitely renormalized in higher orders in
perturbation theory and thus the experimental constraint $\bar \theta
\leq 10^{-9}$ remains a mystery (although it should be mentioned that
the first finite contribution to $\bar \theta$ appears separately form
the infinite part and is small) \cite{ellis}.

\end{enumerate}

\subsection{Dark Matter and The Standard Model}

 Here unfortunately there is basically nothing to be said. By definition
neutrino is massless in the Standard Model (see below, though) and there
are no other light enough particles to give us candidates for the Dark
Matter. Thus, if we could be sure that there is non baryonic DM
in the universe, this would be a clear signal for the new physics.

\subsection{Solar Neutrino Puzzle}

 The solar neutrino puzzle  (SNP) has been discussed extensively by
 Aleksei Smirnov. He did not discuss the magnetic moment solution to
 SNP, and so I will skip it too. We are left then with either
\begin{enumerate}
\item {\bf MSW mechanism}. This, by far the most
popular and elegant solution to SNP is based on the resonant neutrino
oscillations in matter. It favors  the neutrino mass difference and the
mixing angle
\begin{eqnarray}
M S W :  \; \; & \Delta m^2 & \simeq  10^{-7} - 10^{-5} \, eV^2   \nonumber \\
  & \sin^2 2\theta & \simeq  10^{-3} - 10^{-2} \; \;  (small \;  mixing)
\nonumber \\
    & \sin^2 2\theta & \simeq  0.6 - 0.85 \; \;  (large \; mixing) \nonumber
\end{eqnarray}

or

\item {\bf Just-so solution}. This is the long oscillation
length solution to
SNP which corresponds to $\nu_e$ turning into $\nu_\mu$ or $\nu_\tau$
as it arrives to Earth from the Sun. In this case the same parameters
become

\begin{eqnarray}
Just-So :  \; \; & \Delta m^2 &\simeq  10^{-11} - 10^{-10} \, eV^2
 \nonumber\\
 & \sin^2 2\theta  &\geq 0.8 \nonumber
\end{eqnarray}

\end{enumerate}

In what follows let me make the usual assumption that there is a solar
neutrino puzzle and that it is a problem of neutrino physics. As I
said neutrinos are massless and also they carry no magnetic moment
according to the standard picture, hence the Standard Model can offer
no light on SNP. Right? Wrong. Well, maybe wrong. Namely, the question
is whether we should include gravity as a part of the Standard Model;
it certainly would not be unreasonable to do so. Now, we say that
neutrinos are massless in the Standard Model, since the lepton number
is conserved and since by definition there are no right handed
neutrinos. In other words, an effective interaction term
\begin{equation}
\nu_L^T C \nu_L {(\phi^0)^2 \over M}
\label{int-term}
\end{equation}
($\phi^0$ is the neutral component of the Higgs doublet) can arise in
higher orders only if B-L symmetry is
broken ($M$ would be the scale of B-L breaking).  The essential point
here is that gravity through its non-perturbative Planck scale effects
(virtual black holes, virtual wormholes, virtual... holes) may break
B-L, which, as we know, is an accidental anomaly-free global symmetry
in the Standard Model. Thus (\ref{int-term}) could be generated with
$M = M_{Pl}$, giving us
\begin{equation}
m_{\nu} = \frac{\langle  \phi \rangle^2}{M_{Pl}} .
\label{numass}
\end{equation}
In other words, neutrino becomes massive with $m_{\nu}$ approximately
$10^{-6} - 10^{-5} \, eV$. And if $\Delta m_{\nu}^2 \simeq m_{\nu}^2
\simeq 10^{-12} - 10^{-10} \, eV^2$, and if the neutrino mixing angles are
large (a democratic acting of gravity?), this would suffice to provide
the large oscillation length solution to the SNP \cite{gravitySNP}.
This was discussed in my lectures.

\subsection{Baryogenesis}

Again the Standard Model offers an unexpected richness in providing all
the ingredients needed for the creation of baryon asymmetry.  The crucial
 point is the anomaly induced quantum
mechanical breaking of the baryon number, otherwise classically
conserved. The predicted baryon asymmetry $n_B /
n_{\gamma}$ is still subjected to an intense debate. It depends
strongly on the nature of the phase transition at  temperatures of
order $M_W$, and it may be some time before we have a definite answer
to this important, if not central, issue of the Standard Model.
Furthermore, it is not clear if there is enough CP violation built in
the Kobayashi-Maskawa mechanism to provide a large enough $n_B /
n_{\gamma}$.  I shall not go into details here for this  was not
discussed at the School (for a review, see \cite{misha}).

Let me stress though an important property of $n_B / n_{\gamma}$ that
could come out of the Standard Model. Since the CP-violation at $T=0$
and at $T\leq M_W$ originate from the same source, you can in principle
predict the sign of baryon asymmetry. By contrast, conventional GUTs
fail in this respect since the new interactions of superheavy Higgs and
gauge bosons contain new CP-phases, and there is no one-to-one
correspondence between the signs of CP-violation at high and zero
temperature. For example, the superheavy $X$ bosons of $SU(5)$ with
charges $- 4/3$ have the following interactions \cite{moha}
\begin{equation}
L_{int}(X)= g  X^\mu [ \bar P_L^c \gamma_\mu K P_L]
\end{equation}
where

\begin{equation}
P = \left( \begin{array}{c}  u\\c\\t\end{array} \right) \; ; \; K =
\left( \begin{array}{ccc}  e^{i\phi_u} & & \\ & e^{i\phi_c} & \\
 & & e^{i\phi_t}\end{array} \right)
\end{equation}
in the basis where quark masses are positive numbers. The matrix K is
completely independent of the Kobayashi-Maskawa matrix, thus new
phases. The same is true of other gauge boson and superheavy Higgs
interactions.

It is clear that even in the Standard Model there may be rich
implications for cosmology. However, as we have seen, the Dark
Matter chapter is rather dull, and since the School was mainly
devoted to Dark Matter, it is not surprising that we will go
beyond {\em un poquito}.

\section{Beyond the Standard Model}

\subsection{Supersymmetry}

In recent years, Supersymmetry has grown to be the standard extension of
the Standard Model. The reason is not just the structural beauty of the
theory, but rather its role in providing a mechanism to keep the weak
scale so many orders of magnitude below the Planck and GUT scales. A
renormalizable theory such as the Standard Model should not know of
large scales, but unfortunately the Higgs mass is sensitive to any new
scale above $M_W$. For this reason, the Standard Model is crying for
supersymmetry.

We can say that supersymmetry for Higgs mass plays the same role as
chiral symmetry does for the fermion mass. A chiral symmetry makes $m_f
= 0$ natural, and if we have supersymmetry clearly the vanishing of the
scalar mass becomes natural too.  For this to work in the real world,
the scale of supersymmetry breaking should be of order $M_W$. This was
emphasized in Graham Ross's lectures at the School. He has discussed
extensively the phenomenology and cosmology of supersymmetry, and I
refer you to his lecture notes. It is hard to overstress the richness
of the supersymmetric phenomenology. We are to find the whole replica
of our world with every fermion interchanged by a boson and viceversa
at the energy range $1 -10 \, TeV$.

In the minimal supersymmetric Standard Model, the superpotential
 assuming matter parity is

\begin{equation}
W = \mu H \bar H + g_u H Q U^c + g_d \bar H Q D^c + g_e \bar H L E^c
\end{equation}

\vspace{0.5cm}
where  $H$ and $\bar H$ are the  necessary two Higgs doublet
superfields with opposite hypercharges, and $Q,U^c, D^c, L, E^c$ are
quark and lepton superfields respectively; $g_{u,d,e}$ are ``Yukawa''
coupling constants (family indices are suppressed) and $\mu$ is a mass
parameter which has to be of order $m_W$ for the correct electroweak
symmetry breaking. It has a continuous global R-symmetry under which the
respective charges are $R_H=R_{\bar{H}}=1$, and $1/2$ for the rest of
the superfields. Assuming soft-breaking terms coming from the hidden
sector breaking of supergravity leads to the following expression for
the potential

\begin{eqnarray}
V =\sum_i|\frac{\partial W}{\partial \phi_i}|^2 &+& m_gAW^{(3)}+
m_gBW^{(2)}+ h.c. \nonumber \\
& &+ \sum_i m_g^2|\phi_i|^2+(D-terms)
\end{eqnarray}

where $\phi_i$ are the scalar components of the superfields and
$W^{(2)}$ and $W^{(3)}$ are its bilinear and trilinear (in $\phi_i$)
pieces respectively,   $A$ and $B$ are  numbers related to the
details of the hidden sector, and $m_g$ is the gravitino mass. The $A$
and $B$ pieces  clearly break R-symmetry.

Notice that you still end up with a discrete R-parity symmetry under
which $H$ and $\bar H$ are invariant  and all others superfields (and
the superfield variable $\theta$) change the sign. This is equivalent
to all usual particles being invariant and all supersymmetric partners
being odd under R-parity. Thus the lightest supersymmetric particle
(LSP), normally expected to be one of  the neutralinos, is stable. This
is true in any model where R-parity is not broken, i.e. in the models
where the sleptons do not get a non zero vev

The important role of Supersymmetry in providing the Dark Matter of the
Universe was discussed by Antonio Masiero. From what we said, the LSP
becomes naturally a Dark Matter candidate. For details see Masiero's
lectures.

I do wish to emphasize one important property of supersymmetric Grand
Unification. We said that Supersymmetry is tailor-made for the Standard
Model Higgs mass problem; what comes as a bonus is that it also
provides a way out of  the proton lifetime problem of minimal Grand
Unification. Namely, in minimal $ SU(5)$ the proton is too short-lived
and furthermore $\sin^2\theta_W$ comes out to be too small. On the
other hand the presence of the supersymmetric partners, if their masses
are of order $M_W$, increases proton lifetime and  in the minimal
theory predicts correctly $\sin^2\theta_W = 0.23$.

Another interesting aspect of Supersymmetry are the restrictions
on the Higgs potential. With the most general soft-breaking term, the
 potential takes the form

\begin{eqnarray}
V  = m_1^2 \Phi^{\dag} \Phi +  m_2^2 \bar \Phi^{\dag} \bar \Phi &+&
 (m_3^2 \bar \Phi i \tau_2 \Phi + h.c. )   \nonumber \\
& &+ \frac{g^2 + g'^2}{8} ( \Phi^{\dag}   \Phi - \bar \Phi^{\dag}
\bar \Phi )^2
\end{eqnarray}

where $\Phi$ and $\bar \Phi$ are the Higgs doublets from $H$ and $\bar
H$ superfields. It is not hard to see that we have an important
prediction of the lightest Higgs particle being less than the Z boson.
Unfortunately, this is somewhat obscured by the radiative corrections,
as has been discussed by Rosiek.

Finally, the question is what happens to baryogenesis at the weak scale
in the context of Supersymmetry. At the first glance the situation is
analogous to the Standard Model one, but there is more to it. We heard
from Massimo Pietroni that the high temperature behavior of
supersymmetric theories is rather interesting; he and  his
collaborators find that there may be more CP-phases generated at high
temperature. This may have important implications for baryogenesis.
However, again this prevents one from computing the sign of baryon
asymmetry.

Before going on, I would like to remind you that we still lack {\em
the} explanation of the splitting of the large and the small scales
(the so-called doublet-triplet splitting problem). For some interesting
ideas in this direction, see \cite{dvali}.

It is clear that in this short review I cannot do justice to such an
important subject as Supersymmetry. I refer you to Ross's lectures and
also to Ref. \cite{supersymmetry}

\subsection{Neutrino Physics and Cosmology}

It is well known that there are almost as many neutrinos in the
universe as there are photons, approximately $150$ per cubic centimeter
per species. Whether or not this neutrino sea determines the fate of the
Universe depends clearly on neutrino mass. The mechanisms to generate
appreciable neutrino mass have been discussed by me and in greater
detail by Jose Valle. The principal framework to understand the
smallness of neutrino mass lies in the see-saw mechanism
\cite{see-saw}, which as is well known,  leads to a  neutrino mass
strongly suppressed compared to the  Dirac mass terms, $m_D$
\begin{equation}
 m_{\nu} \simeq  m_D {m_D \over M_R} \label{see-saw}
\end{equation}
 where $M_R$ is the mass of the right-handed neutrino
(assume $M_R\geq M_W$).  The crucial question we need to answer in
order to make predictions is what is the value of $M_R$. We have
mentioned before that $M_R = M_{Pl}$ is quite plausible; unfortunately
except for the potential solution of the SNP nothing interesting would
emerge in this case.

Some possible natural scales for $M_R$:
\begin{itemize}
\item $M_R= M_X \simeq 10^{16} \, GeV$ (GUT scale)\\
Then,
\begin{equation}
m(\nu_{\tau}) \simeq 10^{-3} \, eV
\end{equation}
 and other
neutrinos much lighter. Obviously, this is too small to give us
enough Dark Matter, but fits ideally with the values needed for
 the MSW solution of the solar neutrino puzzle.

\item $M_R = 10^{11}-10^{12}  \, GeV$ (an intermediate scale in $SO(10)$)\\
Now ,
\begin{equation}
m(\nu_{\tau}) \simeq  1 -10 \, eV ,
\end{equation}
ideally fitting the Dark Matter constraint.The MSW mechanism
would then go through the lighter neutrinos mass splitting.

\item $M_R \simeq M_W$ (electroweak see-saw)\\
The interesting feature of the electroweak see-saw is that the electron
neutrino could be the Dark Matter of the Universe, and at the same time
could provide an observable neutrinoless double beta decay (but no
SNP). You could also generate $M_R$ through a vev of some singlet
scalar $S_R$ which leads to the existence of a Majoron $G$, a Goldstone
boson of B-L symmetry breaking \cite{majoron}. For $M_R\simeq M_W$ the
Majoron can be an important Higgs decay product since Higgs can now
decay into two Majorons. This was emphasized by  Valle and also by
Romao. The heavy neutrinos can decay into $\nu_e$ an the Majoron, in
accord with all the cosmological constraints.  \end{itemize}

Another important mechanism of neutrino masses is their radiative
generation, which was discussed by Valle. The simplest prototype model
which implements this mechanism is  that due to Zee,  which consists in
adding another Higgs doublet and a singlet charged field $h^+$ to the
Standard Model. The B-L symmetry is broken either explicitly or
spontaneously as above. Such models have a richer phenomenology than
the see-saw ones (new Higgs scalars) and also the resulting neutrino
masses and mixing angles have interesting patterns with some general
potentially identifiable characters \cite{alioshatao}.

The Majoron idea can be naturally implemented in supersymmetry; all one
needs to do is to break the global R-symmetry spontaneously. Models of
this kind offer a rather rich phenomenology and have been covered in
great detail by Romao and Valle.

Their important characteristic is that they can provide both Dark
 Matter and the MSW solution to the SNP.

As can be clear even from our brief discussion, taking SNP or DM
constraints seriously leaves us still with a lot of freedom in the
neutrino mass matrix. Suppose now that you take seriously both, and
furthermore let us also take seriously the problem of atmospheric
neutrinos. AN has been summarized nicely by Evgueni Akhmedov and
basically it says that we observe about a half of  the predicted ratio of
muon to electron
neutrinos in the atmospheric neutrinos. The simplest explanation could
be oscillations of $\nu_\mu$ into $\nu_e$ or $\nu_\tau$ or even some
new sterile neutrino. The relevant mass difference and mixing angle
should be

\begin{eqnarray}
AN :  \; \; & \Delta m^2 & \simeq 10^{-3} - 10^{-1} \, eV^2  \nonumber\\
 & \sin^2 2\theta  & \simeq 0.4  - 0.6 \; \; (\nu_\mu \to \nu_\tau) \nonumber
\\
& \sin^2 2\theta  & \simeq 0.3  - 0.8 \; \;  (\nu_\mu \to \nu_e) \nonumber
\end{eqnarray}

If you insist on the minimal neutrino spectrum of only three $\nu$'s,
you are forced to assume an almost degenerate mass matrix, with a
typical eigenvalue of $1 \, eV$. Namely this is required by the DM
constraint and the mass differences must be small in order to explain
SNP and AN. This was discussed by Daniele Tomasini.

I find it rather interesting that a realistic possibility can follow
from the see-saw mechanism together with some family symmetry
\cite{degenerate}. Namely, in the simplest versions of $SO(10)$ or
left-right symmetric models, there is a direct mass term for a
left-handed neutrino
\begin{equation}
m(\nu_L) \simeq \alpha {\langle \phi \rangle^2 \over M_R}
\label{mnul}
\end{equation}

where $\alpha $ is some dimensionless constant \cite{mohagoran}.
Augmented with a horizontal symmetry, this would imply a diagonal,
degenerate neutrino mass matrix. In this picture the mass splittings
and non-vanishing mixing angles have their origin in the usual see-saw
contribution \\ $m_D^2 / M_R$.

On the other hand, it could be that there are additional light
neutrinos (which must be sterile in order no to upset Z decays), as has
been advocated strongly by Juha Peltoniemi. In this case it is not
surprising that one can satisfy all of the above constraints without
the necessity of a degenerate mass matrix. In general, one does not
expect sterile neutrinos to be light, but Enrico Nardi showed us an
ingenious example based on $E_6$ GUT which predicts naturally light
sterile neutrinos.

\subsection{Strong CP Problem and Axions}

We said before that the strong CP problem is one of the central
unresolved issues in the Standard Model. There are basically two
different ways of resolving it.

 \begin{enumerate}

 \item {\bf Dynamical
relaxation or Peccei-Quinn mechanism}. The most elegant and the most
popular solution to the strong CP problem still remains the
Peccei-Quinn mechanism. By adding another doublet to the standard
model, we introduce a $U(1)_{PQ}$ symmetry whose spontaneous breaking
gives an axion, the would-have-been Goldstone boson if not for the
instanton effects which give it a small mass. The astrophysical
constraints imply that the scale $M_{PQ}$ of $U(1)_{PQ}$ must satisfy $
M_{PQ} \geq 10^9 \, GeV$, and if $M_{PQ}$ is a couple of orders of
magnitude larger the axion could be the Dark Matter of the Universe.

Since all the couplings of the axion to the matter and light are
inversely proportional to the scale $M_{PQ}$, it is not surprising that
the axion has not been found yet. Both the cosmological implications
and the search for the axion were described in detail by Turner.

Although  a beautiful mechanism, the axion picture suffers from the ad
hoc way in which it is introduced. It would be much nicer if
$U(1)_{PQ}$ were an automatic consequence of some more fundamental
symmetry such as for example the family or the GUT symmetry. There may
be more to this than esthetics, since this could be a way out of
potentially catastrophic gravitational effects \cite{blackholes}.

\item {\bf Symmetry principle: P or/and T}. In this approach, the
strong CP parameter $\bar \theta$ becomes a finite and calculable
quantity in perturbation theory \cite{PorT}. For example parity or
left-right symmetry would imply
\begin{equation}
M_q^{\dag} = M_q
\end{equation}
and the departure from the hermiticity can be shown to be finite.

Although both P and T are arguably more natural symmetries than a
global $U(1)_{PQ}$ broken by instantons, the resulting models lack the
simplicity and the beauty of the PQ mechanism.

\end{enumerate}

\section{Summary and Outlook}

As I have tried to convey in these few pages, the Standard Model of
electroweak interactions is in excellent shape. Its only fault may be
the lack of the  excitement of new predictions. This is especially true
as far as cosmology is concerned. We have no DM candidate
whatsoever, we have only a little hope of explaining the SNP. As far as
baryogenesis is concerned, we must wait more before we will have a
clear picture.

What can we expect from the new physics beyond the Standard Model?
 Let
us summarize here the most interesting  potential cosmological
consequences.

\subsection{Dark Matter}
\begin{itemize} \item {\bf Neutrino}. Clearly, since we know that
 they
exist and fill up the universe, neutrinos are the most natural
candidates for DM. Assuming that one of their masses lies in the
electonvolt region, they would provide what is called a Hot Dark
Matter, HDM (they are relativistic at the time of decoupling).
Interestingly enough, any of $\nu_e, \nu_\mu$ or $\nu_\tau$ could
 be Dark Matter, if not all three of them.

It is no problem to come up with models that give the desired
value of neutrino mass, the trouble is the opposite: there are
 too many such models. We desperately need more information from
solar  and atmospheric neutrinos before we can
distinguish between the many theories of neutrino mass. We have
seen that if all the puzzles remain to be true, the neutrino
 mass matrix would be known to a good precision.

\item {\bf Axion}. Unlike  neutrinos,  axions are  hypothetical
particles. It is true though that their existence is theoretically
well founded and, what is even more important, if they do exist
they are likely to be the Cold Dark Matter  (CDM) whose existence
seems to be required from the recent COBE findings. Namely, unlike
neutrino mass which is a free parameter with only  an upper bound,
the axion mass (or in other words, $M_{PQ}$) has to lie close to the
desired  CDM value. However, if the axion is not to be found we
 will have to consider the other candidate solutions of the
 strong CP problem.

\item {\bf Neutralinos}.
 The lightest of them (LSP), if stable, is clearly a natural candidate
  for the Dark Matter.

\end{itemize}

The ideal Dark Matter scenario after the COBE findings appears to
 be some percentage of HDM and the rest mostly CDM. My prejudice
is that these roles are played by neutrinos an the axion. Of
 course there could be baryonic Dark Matter in the Universe in
the form of MACHOs which were discussed by de R\'ujula and Masso.

\subsection{Solar Neutrino Puzzle}
It may not be certain that there is a Solar Neutrino Puzzle or
that it
may be related to neutrino physics. However, the most popular
theoretical prejudice is that neutrinos are massive and thus mix,
 which makes them naturally oscillate into each other. If
so, we have a beautiful MSW mechanism which is still {\em the} most
natural explanation of SNP.

Although less elegant, the idea of long oscillation length (just-so)
solution of SNP may result naturally from non-perturbative
gravitational effects. In this case {\em we may not have to modify the
Standard Model at all}.

\subsection{Baryogenesis}

Here we have a rather exciting situation. It is still not clear that
the Standard Model itself cannot produce enough baryon asymmetry, the
same can be said of its minimal supersymmetric extension. I shall not
go into the controversy surrounding this issue.

Of course if  it is necessary to extend the Standard Model it is easy
to produce enough baryon asymmetry. For example one may simply enlarge
the Higgs sector to two doublets and introduce new sources of CP
violation. Or one may resort to the good old GUTs and produce enough B$-$L
asymmetry which cannot be erased by the breaking of B$+$L symmetry trough
the anomaly. The problem is then that you loose the connection between
the $K-\bar K$ system, and the baryon-number violating part of the
theory, i.e. one cannot predict the sign of the asymmetry itself.

I would also like to add that it would be really nice to come up with a
theory that predicts both the value and the sign of baryon asymmetry
and simultaneously the amount of strong CP breaking.

\subsection{Outlook}
In summary, what can we expect in the years to come ?  I would say that
it is extremely likely that there is another Higgs doublet, for as we
have seen almost any extension of the Standard Model points to it. This
tells you that i believe that there is an elementary Higgs scalar
to start with.
Also, I for one would not be surprised if we discover a new world of
supersymmetric partners. I am convinced that neutrino has a mass
although what its value is  is far from clear. My prejudice, based on
cosmology, is that at least one of the neutrinos has a mass in the $eV$
region as to give us the Hot Dark Matter. And if the Cold Dark Matter
is necessary as it appears, I believe that it will be in the form of
axions.

So much about my prejudices. I hope the future will prove me completely
wrong. Notice however that we may not have to change the gauge
structure of the Standard Model at all. As we have seen, all the
essential extensions of the Standard Model  such as Supersymmetry,
Peccei-Quinn symmetry and  those that provide neutrino mass and
enough baryon asymmetry may all be constructed within the $SU(2)\otimes
U(1)$ gauge theory.

\subsection*{Acknowledgements}

I have not only enjoyed this excellent meeting and the beautiful city
of Valencia;
it is here that I broke the ice and started to speak spanish. Quisiera
expresar mi agradecimiento a los organizadores  de la Escuela,
 especialmente a Fernando P\'erez y Jose Valle por su hospitalidad
y por haberme dado la oportunidad de ver los mejores fuegos
artificiales del mundo.

\end{document}